\documentclass[apj]{emulateapj}

\newcommand{\eg}{e.g., }
\newcommand{\ie}{i.e., }
\newcommand{\Msun}{M_{\odot}}
\newcommand{\kms}{km~s$^{-1}$}

\newcommand{\Cofs}{$^{56}$Co}
\newcommand{\Nifs}{$^{56}$Ni}

\def\gsim{\mathrel{\rlap{\lower 4pt \hbox{\hskip 1pt $\sim$}}\raise 1pt
\hbox {$>$}}}
\def\lsim{\mathrel{\rlap{\lower 4pt \hbox{\hskip 1pt $\sim$}}\raise 1pt
\hbox {$<$}}}

\def\ion#1#2{{\rm #1}~{\sc #2}}

\shorttitle{Spectropolarimetry of SN 2005bf: Larger Asymmetry}
\shortauthors{Tanaka et al.}
 
\begin{document}

\title{
Spectropolarimetry of the Unique Type I\lowercase{b} Supernova 2005\lowercase{bf}: Larger Asymmetry Revealed by Later-Phase Data
\altaffilmark{1}}
\author{
Masaomi Tanaka\altaffilmark{2,3}, 
Koji S. Kawabata\altaffilmark{4},
Keiichi Maeda\altaffilmark{3},
Masanori Iye\altaffilmark{5},
Takashi Hattori\altaffilmark{6},
Elena Pian\altaffilmark{7},
Ken'ichi Nomoto\altaffilmark{3,2},
Paolo A. Mazzali\altaffilmark{8,9}, and
Nozomu Tominaga\altaffilmark{10,5}
}

\altaffiltext{1}{Based on data collected at Subaru Telescope, 
which is operated by the National Astronomical Observatory of Japan.}
\altaffiltext{2}{Department of Astronomy, Graduate School of Science, University of Tokyo, Bunkyo-ku, Tokyo, Japan; mtanaka@astron.s.u-tokyo.ac.jp}
\altaffiltext{3}{Institute for the Physics and Mathematics of the Universe, University of Tokyo, Kashiwa, Japan}
\altaffiltext{4}{Hiroshima Astrophysical Science Center, Hiroshima University, Higashi-Hiroshima, Hiroshima, Japan}
\altaffiltext{5}{Optical and Infrared Astronomy Division, National Astronomical Observatory, Mitaka, Tokyo, Japan}
\altaffiltext{6}{Subaru Telescope, National Astronomical Observatory of Japan, Hilo, HI}
\altaffiltext{7}{Istituto Naz. di Astrofisica-Oss. Astron., Via Tiepolo, 11, 34131 Trieste, Italy}
\altaffiltext{8}{Max-Planck Institut f\"ur Astrophysik, Karl-Schwarzschild-Strasse 2 D-85748 Garching bei M\"unchen, Germany}
\altaffiltext{9}{Istituto Naz. di Astrofisica-Oss. Astron., vicolo dell'Osservatorio, 5, 35122 Padova, Italy}
\altaffiltext{10}{Department of Physics, Konan University, Okamoto, Kobe, Japan}

\begin{abstract}
We present an optical spectropolarimetric observation of the unique Type Ib
supernova (SN) 2005bf at 8 days after the second maximum.
The data, combined with the polarization spectrum taken at 6 days before 
the second maximum (Maund et al. 2007a), enable us to closely examine 
the intrinsic properties of the SN.
The polarization percentage is smaller at the later epoch
over a wide wavelength range,
while the position angle is similar at the two epochs.
We find that an assumption of complete 
depolarization of strong lines at the emission peak
is not necessarily correct.
The intrinsic polarization of the SN is larger,
and thus, the ejecta of SN 2005bf would be more asymmetric 
than previously expected. 
The axis ratio of the photosphere projected on the sky deviates 
from unity by at least 20 \%.
If the position angle of interstellar polarization is aligned with 
the spiral arm of the host galaxy, the deviation is larger than 25 \%.
The line polarization at the \ion{He}{i}, \ion{Ca}{ii} 
and \ion{Fe}{ii} lines is also smaller at the later epoch.
The changes in the position angle across these lines, 
which were observed at the earlier epoch, are still marginally present 
at the later epoch.
The similar polarimetric behavior suggests that the distributions of 
these ions are correlated.
Properties of polarization, as well as the light curve and the spectra 
both at photospheric and nebular phases,
can be explained by an aspherical, possibly unipolar explosion
of a WN star in which the blob of \Nifs\ penetrates C+O core 
and stops within the He layer.
\end{abstract}


\section{Introduction}
\label{sec:intro}

Asymmetry is one of the keys to understand 
the explosion mechanism of core-collapse supernovae (SNe).
Various mechanisms that could lead to aspherical explosions
have been studied, including \eg rotation and magnetic field 
\citep[\eg][]{mue81,yam94,tak04}, 
convection \citep[\eg][]{bur95,jan96} or
standing accretion shock instability \citep[\eg][]{blo03,iwa08}.

Observational constraints on SN asymmetry can be obtained 
via direct imaging \citep[\eg][]{wan02,hwa04}
only for a few, very nearby SNe or supernova remnants.
For extragalactic SNe, polarimetry is one of the most direct 
methods to study asymmetry of the explosion.
Polarization is produced by electron scattering within the SN ejecta.
Since the polarization vectors are canceled out 
in spherically symmetric ejecta, 
no polarization would be detected from a spherical explosion.
In other words, the detection of polarization undoubtedly indicates 
asymmetry of the explosion \citep{sha82,mcc84,hof91}.

Spectropolarimetry is a more powerful tool than imaging polarimetry 
because polarization across lines 
possesses information on the element distribution
\citep[see][for a recent review]{wan08}.
Since the scattering by the line depolarizes the light,
line polarization is detected when the distributions
of elements or ions are not uniform
(even when the underlying photosphere is spherically symmetric).
The asymmetric nature of core-collapse SNe has 
been studied in detail by spectropolarimetry
\citep[\eg][]{cro88,tra93,wan01,wan03a,leo01,leo02,leo06,kaw02,kaw03,mau07a,mau07b,mau07c,hof08,tan08a}.

In this paper, we present a spectropolarimetric observation of the Type 
Ib SN2005bf, which appears to be a unique supernova because of the 
following aspects:
(1) The light curves showed double peaks 
\citep{anu05,tom05,fol06}. 
(2) The SN was at first classified as Type Ic (SNe without H or He lines) 
but then re-classified as Type Ib (SNe with He lines) 
because of the emergence of strong He lines \citep{wan05,mod05}.
(3) The decline of the luminosity after the second maximum was 
very rapid \citep{tom05,fol06}.
In addition, the luminosity around 300 days after the explosion
was much fainter than expected from $0.3 \Msun$ of \Nifs\
necessary to account for the second SN luminosity maximum \citep{mae07}.

In order to explain the two peaks of the light curve
by radioactivity, the distribution of 
\Nifs\ must have two components.
This fact led 
\citet{tom05} and \citet{fol06} to propose a jet-like explosion
for SN 2005bf.
However, the low luminosity at nebular phases may suggest
another heating source for the second peak,
such as central remnant's activity \citep{mae07}.

\citet{mau07a} (hereafter M07) presented a spectropolarimetric 
observation of SN 2005bf taken with the Very Large Telescope (VLT)
at 6 days before the second visual maximum,
\ie at $t= -6$ days, where $t$ indicates the epoch 
relative to the second visual maximum, JD 2453498.8 \citep{fol06}.
Based on detected large polarization levels and
changes in the polarization position angle at the lines of \ion{He}{i}, 
\ion{Ca}{ii} and \ion{Fe}{ii},
they suggested a tilted-jet model, \ie a scenario in which 
the asphericity is due to the ejection of blob(s)
(or jets) in a tilted direction with respect to the 
symmetry axis of the SN.

One difficulty in polarimetric observations is accounting for the degree
of interstellar polarization (ISP) caused by interstellar extinction 
\citep{dav51}. Although a correct estimate of ISP is essential 
to determine the intrinsic polarization of the SN, 
such an estimate is especially difficult to determine with an observation 
at a single epoch.

We present spectropolarimetric data of SN 2005bf
taken with the 8.2 m Subaru telescope at 14 days after the observation by M07.
The two-epoch data enable us to estimate ISP more carefully.
In \S \ref{sec:obs}, we present the observations and data reduction.
The data taken at the two epochs are compared, and
updated estimates of ISP are shown in \S \ref{sec:res}.
In \S \ref{sec:dis}, our interpretation of 
spectropolarimetric data is discussed.
Finally, we give the conclusions in \S \ref{sec:con}.

\section{Observations and Data Reduction}
\label{sec:obs}

Spectropolarimetry of SN 2005bf was performed 
with the 8.2 m Subaru telescope equipped with the 
Faint Object Camera and Spectrograph \citep[FOCAS,][]{kas02}
on 2005 May 15 UT (JD 2453506.76).
This epoch corresponds to $t= +8$ days 
(14 days after the observation by M07).

We used a slit of $0.8''$ width and two 300 lines mm$^{-1}$ grisms.
Blue ($\lambda < 6900$ \AA) and red ($\lambda > 6900$ \AA) 
spectra were taken separately.
No filter was used for the blue spectrum,
while the O58 filter was used for the red spectrum
to eliminate the second order light.
The wavelength resolution is $\lambda/\Delta \lambda \sim 650$.

The linear polarimetric module of FOCAS consists of 
a rotating superachromatic half-wave plate 
and a crystal quartz Wollaston prism.
Both the ordinary and extraordinary rays are recorded 
on the CCD simultaneously.
From four integrations at the 
$0^{\circ}, \ 45.^{\circ},\ 22.5^{\circ}$ and $67.5^{\circ}$ positions 
of the half-wave plate, Stokes $Q$ and $U$ were derived as in \citet{tin96}.
The total exposure time of four integrations was 2400 sec 
for each part of the spectrum.

The position angle was calibrated by the observation of 
the strongly polarized star HD 155197 \citep{tur90}.
The wavelength dependence of the optical axis of the half-wave plate
was also corrected using this observation.
Since sky conditions were better during the observation of the blue spectrum, 
the polarization data are binned into 30 \AA\ and 50 \AA\ for 
the blue and red spectra, respectively.
This yields an average signal-to-noise ratio S/N $\sim 330$.
For the degree of polarization $P$, the polarization bias was 
corrected using the results of \citet{pat06}.

The flux spectrum was calibrated using the observation 
of the spectrophotometric standard star Feige34 \citep{oke90}
and further calibrated using previously reported photometry of the SN  
\citep{tom05}.
Telluric absorption lines are removed using 
the spectrum of the standard star.

\begin{figure}
\begin{center}
\includegraphics[scale=0.47]{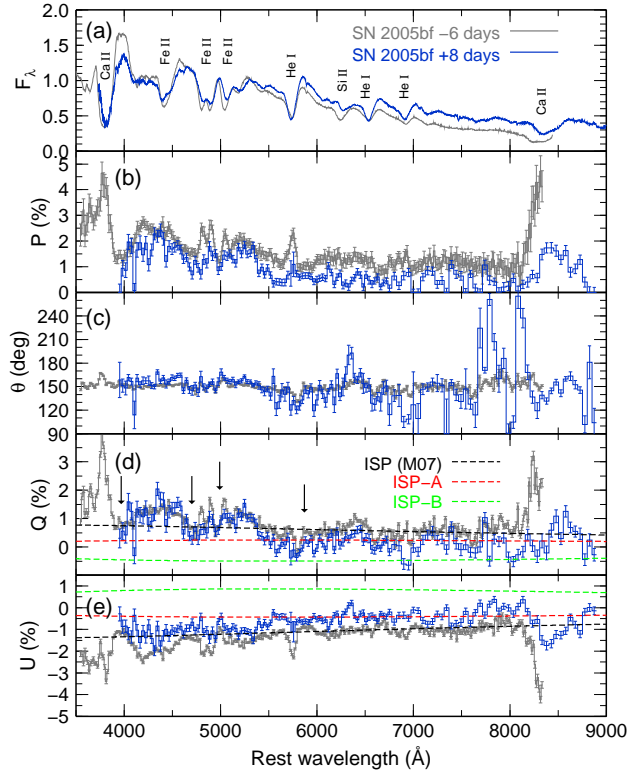}
\caption{
($a$) Spectrum of SN 2005bf 
(blue, $10^{-15} {\rm erg\ s^{-1}\ cm^{-2}}$ \AA$^{-1}$) 
at $t = +8$ days from the second visual maximum 
(JD 2453498.8, \citealt{fol06}) compared with the scaled spectrum at 
$t = -6$ days (gray, M07);
($b$) bias-corrected polarization $P$; ($c$) polarization angle $\theta$;
($d, e$) the Stokes parameters $Q$ and $U$.
$P$, $Q$, $U$, and $\theta$ have {\it not} been corrected for
Interstellar polarization (ISP).
The black dashed line shows the ISP assumed by M07 while
the red and green dashed line shows our updated estimates.
The arrows in panel ($d$) show the position of the 
emission peak of strong lines,
where complete depolarization is assumed.
The data at $t = +8 $ days taken with the Subaru telescope
are binned to 30 \AA\ and 50 \AA\ for the bluer ($\lambda < 6900$\AA) 
and redder ($\lambda > 6900$\AA) parts of the spectrum.
The data at $t =-6$ days taken with VLT (M07) are binned to 15 \AA.
\label{fig:pol}}
\end{center}
\end{figure}

\begin{figure*}
\begin{center}
\includegraphics[scale=0.8]{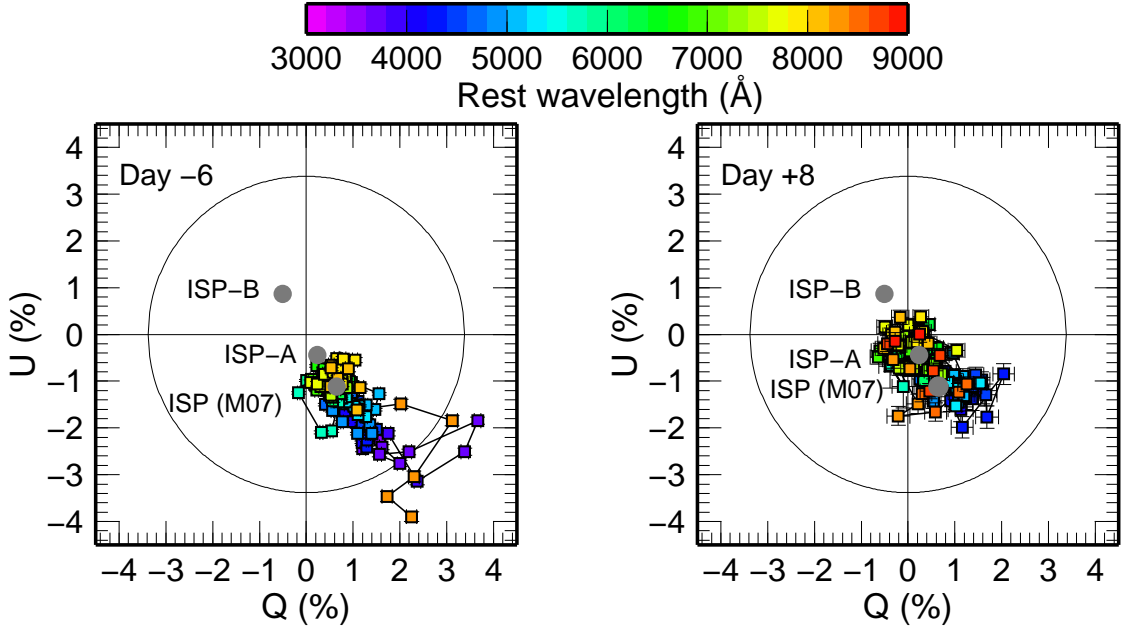}
\caption{
$Q$-$U$ diagram of the polarization data before correction of ISP.
{\it Left}: The VLT data taken at $t =-6$ days by M07.
The data are binned to 30 \AA.
{\it Right}: The Subaru data taken at $t=+8$ days.
The data are binned to 30 \AA\ and 50 \AA\ for the blue and 
red spectra, respectively.
Different colors show the wavelength according to the color scale bar.
The ISPs at 5500 \AA\ estimated by M07 and by this work 
are marked with gray circles.
The large circle shows the upper limit for the ISP ($3.4 \%$).
\label{fig:qu}}
\end{center}
\end{figure*}

\section{Results}
\label{sec:res}

\subsection{Spectropolarimetric Properties at the Two Epochs}

Figure \ref{fig:pol} shows comparison of 
flux spectrum and polarization spectrum
at $t=+8$ days (blue) with those at $t=-6$ days (gray).
Line identifications are given in the panel
for the total light spectrum ($a$).
It is interesting to note that He line velocities increase with time
\citep{tom05}.
This is in contrast to the usual behavior, \ie the photosphere 
recedes and line velocities decrease with time in expanding ejecta.
This peculiar behavior of SN 2005bf may suggest increasing non-thermal 
excitation of the He lines \citep{har87,luc91}.

The Stokes $Q$ parameter at $t=+8$ days is smaller than that at $t=-6$ days 
in the continuum ($d$). 
The difference in the \ion{Ca}{ii} triplet region is larger than
in the continuum.
The difference in the Stokes $U$ parameter between the two epochs 
is more prominent than $Q$ over the whole wavelength range ($e$). 
The $U$ parameter at $t=+8$ days is larger than that at $t=-6$ days
(about half in absolute value).

The difference in the Stokes $Q$ and $U$ between the two epochs
is more easily seen in the $Q$-$U$ plane (Fig. \ref{fig:qu}).
In the $Q$-$U$ plane, the data at $t=+8$ days are closer 
to zero point.
The position angle is similar at the two epochs, $\theta \sim 150^{\circ}$ 
[$\theta \equiv 0.5 {\rm atan}(U/Q)$, see also panel 
($c$) of Fig. \ref{fig:pol}].

As a result, the degree of polarization [$P \equiv (Q^2 + U^2)^{1/2}$]
is smaller at $t=+8$ days over the whole wavelength range
(panel ($b$) of Fig. \ref{fig:pol}).
In the line-free region, the difference in the polarization is $\sim 0.5 \%$.
The polarization at the strong lines,
\eg \ion{He}{i} $\lambda$5876, \ion{Ca}{ii} IR triplet and \ion{Fe}{ii} lines
at 4300-5100 \AA\ is also smaller at $t=+8$ days.
In particular, the polarization in the \ion{Ca}{ii} triplet region 
decreases by a factor of $>2$ in absolute value.

Multi-epoch spectropolarimetry of Type Ib/c SNe 
has not been performed for many objects \citep{wan08}.
One of the well-observed example is Type Ic SN 2002ap.
For SN 2002ap, the continuum polarization level increases with time
before the maximum brightness \citep{wan03a}.
After maximum, the polarization decreases or it is almost unchanged
\citep{kaw02,leo02,wan03a}.
For SN 2005bf, polarization decreases in the interval 
from $t=-6$ to $+8$ days.
However, since we don't know the polarization around the maximum, 
the time evolution could possibly be similar to that of SN 2002ap
\footnote{A decreasing trend in polarization is reported for Type Ic 
SN 2006aj associated with X-ray flash 060218 \citep{gor06},
although the observation consists only in imaging polarimetry
[see \citealt{mau07b} for a single-epoch spectropolarimetry of SN 2006aj].}.

\subsection{Updated Estimates of Interstellar Polarization}

To discuss intrinsic properties of SN polarization,
polarization caused by interstellar extinction (ISP) must be corrected.
First, we give an upper limit of ISP by the total amount of 
reddening \citep{ser75}.
In the line of sight to SN 2005bf, 
Galactic extinction is $E(B-V) = 0.01$ mag \citep{sch98}.
M07 estimated the extinction by the host galaxy to be $E(B-V)<0.37$ mag
by using the equivalent width of the \ion{Na}{i} D line and the relation
given by \citet{tur03}.
Thus, the relation of $P /E(B-V) < 9 \%$ \citep{ser75} gives
the maximum degree of ISP $P < 3.4$ \%.
This maximum value is shown by the large circle in Figure \ref{fig:qu}.

The wavelength dependence of ISP is represented by
\begin{equation}
p(\lambda)\ 
=\ p_{\rm max}\exp \left[ 
-K\ {\rm ln}^2 \left( \lambda_{\rm max}/ \lambda \right) \right],
\end{equation}
which is valid for the Milky Way-like dust \citep{ser75}.
Here $\lambda_{\rm max}$ is the wavelength at the peak of ISP, 
$p_{\rm max}$ is the degree of ISP at $\lambda_{\rm max}$,
and $K$ is given as 
$K = 0.01 + 1.66\ \lambda_{\rm max}\ ({\rm \mu m})$ \citep{whi92}.
M07 estimated ISP 
(the black dashed lines in Fig. \ref{fig:pol})
by assuming complete depolarization
at the emission peak of \ion{He}{i}, \ion{Ca}{ii} and \ion{Fe}{ii} lines
(marked by arrows in panel ($d$) of Fig. \ref{fig:pol}).
This is often assumed when ISP must be estimated from a single epoch 
spectropolarimetry.
As a result, ISP was estimated as $p_{\rm max} = 1.6 \pm 0.2$ \%, 
$\lambda_{\rm max} = 3000$ \AA, 
and $\theta_{\rm ISP} = 149.7^{\circ} \pm 4.0 ^{\circ}$.

However, our data show that the assumption of the complete depolarization 
at $t=-6$ days is not necessarily true.
The polarization levels at the emission peak of 
the \ion{Fe}{ii} lines and \ion{He}{i} $\lambda$5876 
are different between $t=-6$ and $+8$ days.
This fact indicates that the contribution of ISP should be smaller,
as already cautioned by M07.

ISP can be estimated more carefully from multi-epoch data.
The position angle is similar at the two epochs.
This fact suggests that the position angle of ISP is 
$\theta \sim 150^{\circ}$ (similar to the position angle of the observed data) 
or $60^{\circ}$ (the opposite side in the $Q$-$U$ plane)
if the SN does not have a complex change of position angle.
If the position angle of ISP were far from the above two values,
the intrinsic SN polarization would have a complex wavelength dependence
and a time dependence.

If we take a position angle of ISP
$\theta_{\rm ISP} = 149.7^{\circ}$ as in M07, 
a new upper limit of ISP can be obtained by fitting 
the polarization at the emission peak of strong lines at $t=+8$ days,
\eg \ion{He}{i} $\lambda$5876, \ion{Ca}{ii} H\&K and IR triplet 
and \ion{Fe}{ii} lines.
Under the conventional assumption of $\lambda_{\rm max}=5500$ \AA,
good simultaneous fits are obtained with $p_{\rm max} = 0.5 \pm 0.1$ \%.
This is shown with the red dashed lines in Figure \ref{fig:pol}
and in the gray circle in Figure \ref{fig:qu}.
We call this upper limit ISP-A.

It should be noted, however, that the position angle of ISP-A
is orthogonal to the spiral arm of the host galaxy (see Fig. 1 of M07).
This is in contradiction to the expectation that ISP is 
aligned with the spiral arm due to the alignment of 
dusts by the magnetic field \citep{sca87,sca93}.
Thus, it is also quite likely that ISP has a position angle 
$\theta_{\rm ISP} \sim 60^{\circ}$, being aligned with the spiral arm.
In addition, this does not cause any complex 
change in the position angle of SN (see above).
In this case, the degree of ISP is only weakly constrained by 
the total extinction, \ie $P < 3.4 \%$.
We take $p_{\rm max}=1 \%$ as an example.
This ISP, hereafter denoted as ISP-B, is shown in the green dashed lines 
in Figure 1 and in the gray circle in Figure 2.
Since the correct value of $p_{\rm max}$ is unclear, it is cautioned 
that the exact value of the intrinsic polarization is also arbitrary.

\begin{figure}
\begin{center}
\includegraphics[scale=0.47]{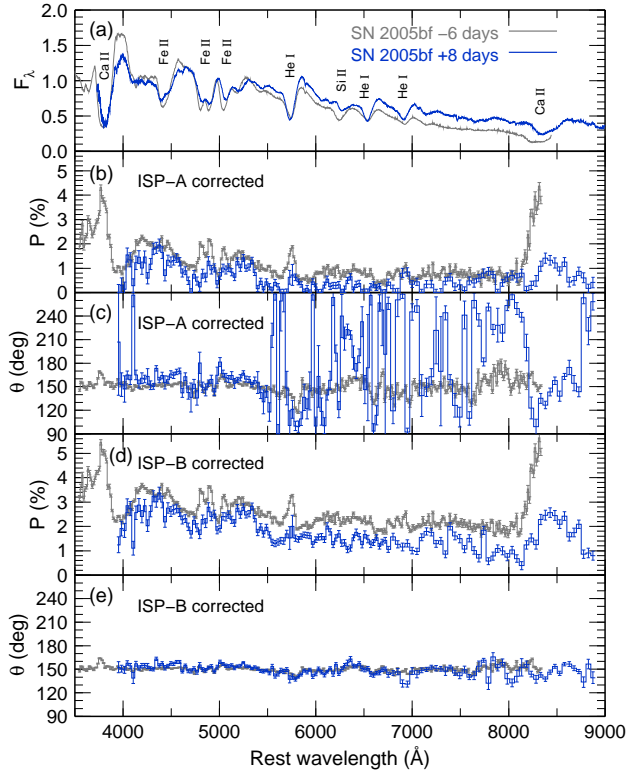}
\caption{
($a$) Spectra of SN 2005bf 
at $t=+8$ days (blue, in $10^{-15} {\rm erg\ s^{-1}\ cm^{-2}}$ \AA$^{-1}$)
and at $t = -6$ days (gray, in scaled flux).
($b,c$) polarization and position angle corrected with ISP-A.
($d,e$) polarization and position angle corrected with ISP-B.
ISP-A is an upper limit of ISP when the position angle is 
$\theta_{\rm ISP} = 149.7 ^{\circ}$
($p_{\rm max}=0.5 \%$ and $\lambda_{\rm max} = 5500$ \AA).
ISP-B is an example case where the position angle of ISP 
is aligned with the spiral arm of the host galaxy: $p_{\rm max}=1.0 \%$, 
$\lambda_{\rm max} = 5500$ \AA, and $\theta_{\rm ISP} = 60^{\circ}$.
\label{fig:ISPpol}}
\end{center}
\end{figure}

\section{Asymmetry of SN 2005\lowercase{bf}}
\label{sec:dis}

Figure \ref{fig:ISPpol} shows 
polarization and position angle corrected with 
ISP-A ($b, c$) and ISP-B ($d, e$).
We discuss asymmetric geometry of SN 2005bf inferred from 
the two-epoch spectropolarimetric data and other observational facts.

\subsection{Continuum Polarization}

Intrinsic continuum polarization 
(defined in the line-free region at 7000-8000 \AA) 
at $t=-6$ days was estimated as $\sim 0.45$ \% 
under the assumption of ISP by M07. 
However, with our updated ISP estimates, it is at least $0.8 \%$
(Fig. \ref{fig:ISPpol}).
Using the results of \citet{hof91}, 
with the opacity at the photosphere $\tau =1$ and oblate geometry, 
this suggests that the axis ratio of the photosphere projected on the sky 
deviates from unity by $\sim$ 20 \%.

If the position angle of ISP is aligned with the spiral arm of 
the host galaxy, intrinsic continuum polarization is 
$\gsim 1.2 \%$.
This suggests even larger asymmetry of the photosphere, 
being $\gsim 25\%$.
Since $p_{\rm max}$ is not strongly constrained in this case, 
the intrinsic polarization is uncertain.
For example, under the assumption of ISP-B ($p_{\rm max}=1\%$),
the axis ratio can be deviated from unity by as large as 50 \%.

After the correction of ISP, 
the difference in the polarization between the two epochs 
is larger in the redder part than in the bluer.
This could indicate that the ISP peaks  
at bluer than 5500 \AA\ as suggested by M07.

The intrinsic polarization level at $t=+8$ days largely depends 
on the choice of ISP.
If the position angle of ISP is $\theta_{\rm ISP} \sim 60^{\circ}$,
the continuum polarization at $t=+8$ days is $>0.5 \%$.
The variation of the position angle along the wavelength is very small
since $\theta_{\rm ISP} = 60^{\circ}$ is
aligned with the observed data in the $Q$-$U$ plane.
In addition, the agreement in the position angle at two epochs is 
extremely good when ISP-B is corrected (panel ($e$) in Fig. \ref{fig:ISPpol}).
This is because ISP-B is out of scatter around the zero point 
in the $Q$-$U$ plane at $t=+8$ days.
These facts may support the possibility that the position angle of ISP
is aligned with the spiral arm of the host galaxy.
Although $p_{\rm max}$ is quite uncertain 
in the case of $\theta_{\rm ISP} = 60^{\circ}$, 
$p_{\rm max}>0.5 \%$ may be preferable 
to avoid a complex wavelength-dependence
of the position angle of the SN at $t=+8$ days.

Since the interval of the two epochs is not close,
the exact evolution of the continuum polarization is not clear.
It is true, however, that
the polarization is smaller at the later epoch.
This can be understood by
(1) decreasing optical depth of electron scattering or 
(2) decreasing asphericity of the photosphere.
In optically thin case, the effect of (1) is important, 
as observed after plateau phase of Type II SNe \citep{leo06}.
However, the epoch of our observation seems to be earlier than 
the epoch when the transition to the nebular phase happens.
Thus, the actual change of the photospheric shape (2) could also
be likely.

\begin{figure}
\begin{center}
\includegraphics[scale=0.8]{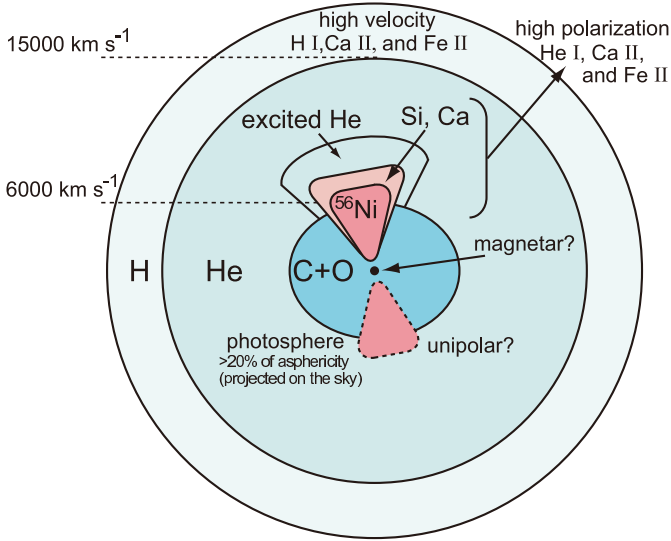}
\caption{
Schematic illustrations of an aspherical explosion model for SN 2005bf.
Our line of sight is near the polar direction.
The progenitor star is a WN star with a thin H layer.
High velocity \ion{Ca}{ii} and \ion{Fe}{ii} lines are formed in the H layer.
The He layer is located at 6,000 - 15,000 \kms.
The blob of \Nifs\ penetrates the C+O core, but not the He layer.
Synthesized, aspherically-distributed \Nifs\ and Ca may be responsible
for the high polarization level of the \ion{Ca}{ii} and \ion{Fe}{ii} lines.
The line forming region of \ion{He}{i} is also aspherical
because of the selective excitation around the \Nifs\ blob.
The photosphere is deformed by the aspherical ejection of the blob 
or jets.
The axis ratio of the photosphere projected on the sky deviates 
from unity by least 20 \%.
The direction where the blob is ejected is tilted with 
respect to the symmetry axis of the photosphere (M07).
The explosion could possibly be unipolar as suggested by the shift
of the nebular emission lines \citep{mae07}.
\label{fig:model}}
\end{center}
\end{figure}

\subsection{Line Polarization}

At both epochs, a large polarization is observed at the 
\ion{He}{i} $\lambda$5876, 
\ion{Ca}{ii} IR triplet and \ion{Fe}{ii} lines.
Little polarization is seen at the \ion{He}{i} $\lambda$6678 
and $\lambda$7065, but this may be due to the weakness of these lines.
The line polarizations at strong lines at $t=+8$ days are smaller 
than at $t=-6$ days.
This can be caused partly by the smaller polarization
of the underlying photospheric radiation at the later epoch.

The difference of the polarization between the two epochs
is the largest at the \ion{Ca}{ii} 
IR triplet region among three strong lines.
The large change may suggest that 
asphericity of the \ion{Ca}{ii} distribution depends 
on the velocity (radius), \ie larger asphericity in the outer layers.
Since the change in the velocities of the \ion{He}{i} and \ion{Fe}{ii} lines
is smaller than that in the \ion{Ca}{ii} line,
the radial dependence of asphericity
cannot be discussed for the He and Fe.

M07 discussed the change in the position angle at the \ion{He}{i},
\ion{Ca}{ii} and \ion{Fe}{ii} lines.
This appears as a loop in the $Q$-$U$ plane and indicates that
the photosphere and the line forming region do not share a
common symmetry axis.
At $t=+8$ days, the change in the position angle at these lines is also seen 
at the \ion{Ca}{ii} line, and marginally 
at the \ion{He}{i} and \ion{Fe}{ii} lines,
although the loop in the $Q$-$U$ plane 
is not as prominent as that at $t=-6$ days.
The similar polarimetric behaviors of these lines, \ie
the similar position angle especially under ISP-B and
the smaller polarization level at the later epoch,
suggest that the distributions of these ions are correlated.

At both epochs, no strong polarization is observed 
at the \ion{Si}{ii} $\lambda 6355$ and the \ion{O}{i} $\lambda 7774$.
Since these lines are very weak, no detection of polarization does
not necessarily indicate sphericity of the distribution of 
\ion{Si}{ii} and \ion{O}{i}.
Although M07 discussed polarization of H lines, 
the H lines are not clearly identified at $t=+8$ days \citep[see][]{fol06}.

\subsection{Aspherical Explosion Model}

We discuss explosion geometry of SN 2005bf.
To explain spectropolarimetric behaviors
as well as the behaviors of the light curve and spectra both at 
photospheric ($\lsim 70$ days after the explosion) and 
nebular ($\sim 1$ year after the explosion) phases,
we suggest an aspherical model as illustrated in Figure \ref{fig:model}.

The progenitor star of SN 2005bf is suggested to be a WN star 
by the identification of the high velocity H$\alpha$ line
\citep{anu05,tom05,fol06}.
The velocity at the bottom of the H layer is $\sim 15,000$ \kms.

The He layer is located below the H layer.
The velocity at the bottom of the He layer is around 6,000 \kms,
constrained by the minimum velocities of the He lines \citep{tom05}.
Since the velocities of the He lines increase with time, 
it is suggested that the excitation of \ion{He}{i} grows with time.
Thus, \Nifs\ is not completely mixed with the He layer.
The distribution of \Nifs\ is clumpy, which is suggested by 
the high polarization level of the Fe lines.
Around the \Nifs\ blob,
non-thermal electrons are created by Compton scattering of $\gamma$-rays
emitted in the decay of \Nifs\ and \Cofs.
Thus, the He lines are formed around the \Nifs\ blob.

Heavy elements such as Si and Ca must also be synthesized around \Nifs.
The polarization of the \ion{Ca}{ii} line that depends on the velocity
(radius, \S 4.2) could be explained by the flared distribution
of synthesized, heavy elements as illustrated in Fig. \ref{fig:model}, 
which is often seen in simulations of jet-like explosion
\citep[\eg][]{tom09}.
Since the absorption of Si is very weak, no detection of 
polarization at the Si line is not inconsistent with this geometry.
The geometry discussed above can explain the similar polarization properties 
of the \ion{He}{i}, \ion{Ca}{ii}, and \ion{Fe}{ii} lines.

The C+O core is located below the He layer at 6,000 \kms.
The photosphere at $t=-6$ and $+8$ days is located in the C+O core 
\citep{tom05}.
The photosphere is deformed so that the projected photosphere 
has $> 20 \%$ asymmetry.
This degree of asphericity can be formed by the ejection of the blob 
(or jets) toward the polar region \citep[\eg][]{kho99,mae02}.
However, the photosphere and the distribution of \Nifs\ do not share 
a common symmetry axis as suggested by the changes in the position angle
across the lines.
This could be due to the fact that the direction 
where the blob is ejected is tilted with respect to the 
symmetry axis of the photosphere 
(as in a tilted-jet model suggested by M07).
Such distributions could possibly be realized by, for example, 
a magnetorotational explosion, where the rotational axis is inclined 
with respect to the magnetic field \citep{mik08}.

The \Nifs\ blob toward us is responsible for the first peak of 
the light curve \citep{mae07}.
It may also explain the velocity shifts of the emission lines seen in 
the nebular spectra.
In this case, the explosion may be unipolar.
In addition, the second peak of the light 
curve is powered by another heating source.
\citet{mae07} proposed magnetar as a central remnant of SN 2005bf
and reproduced the observed light curve until late phases.

\citet{fol06} and \citet{par07} suggested that
the high velocity \ion{Ca}{ii} and \ion{Fe}{ii} lines 
that were present at early epochs ($t \lsim -20$ days)
support the jet-like explosion.
However, since the velocities of these lines coincide to that of H$\alpha$
($v \sim 15,000$ \kms), 
the high velocity \ion{Ca}{ii} and \ion{Fe}{ii} lines are formed 
in the H layer.
If the jet or blob-like structure does not penetrate the He layer
as illustrated in Figure \ref{fig:model}, these lines are unlikely 
to be associated with the blob.
Alternatively, these high velocity lines can result from 
the solar abundance in the H layer because of 
a high electron density in the H layer,
enhancing the recombination of \ion{Ca}{iii} and \ion{Fe}{iii}
\citep{maz05,tan08b}.

\section{Conclusions}
\label{sec:con}

We have presented an optical spectropolarimetric observation 
of the unique Type Ib SN 2005bf taken with the Subaru telescope at $t=+8$ days.
Comparison with the data taken with VLT at $t=-6$ days
enables us to closely study the intrinsic properties of the SN.

Polarization at $t=+8$ days is smaller than that at $t=-6$ days, 
with an almost constant position angle at a wide wavelength range.
We find that 
an assumption of complete depolarization at the emission peak 
of strong lines is not necessarily correct.
This requires reanalysis of the ISP contribution in the observed data.
We put a smaller upper limit for ISP.
Thus, the intrinsic polarization of the SN is larger,
\ie the ejecta of SN 2005bf would be more asymmetric.
The axis ratio of the photosphere projected on the sky 
deviates from unity by at least 20\%.

It is likely that the position angle of ISP is
aligned with the spiral arm of the host galaxy.
If this is the case, the continuum polarization at $t=-6$ days 
is at least $\sim 1.2 \%$.
Then the asymmetry of the projected photosphere is 
even larger, being $\gsim 25 \%$.

The degrees of the line polarization 
at the \ion{He}{i}, \ion{Ca}{ii} and \ion{Fe}{ii} lines similarly
decrease from $t=-6$ to $+8$ days.
The change in the position angle across the line, 
making a loop in the $Q$-$U$ plane, is seen in the \ion{Ca}{ii} line
and also marginally in the \ion{He}{i} and \ion{Fe}{ii} lines
at $t=+8$ days.
The similar behavior of these lines suggests that 
the distribution of these ions are correlated.

We propose an aspherical, possibly unipolar explosion model 
of a WN star as shown in Figure \ref{fig:model}.
In this model, the \Nifs-rich blob penetrates C+O core and stops within 
the He layer.
The direction in which the \Nifs-rich blob is ejected is 
tilted with respect to the symmetry axis of the aspherical photosphere (M07).
The blob is responsible for the first peak of the light curve.
If the explosion is unipolar,
the shifted nebular emission lines are also 
explained by the blob \citep{mae07}.
The non-thermal electrons originating from the \Nifs-rich blob selectively 
excite He around the blob, and this configuration can explain 
the similar polarization properties of 
the \ion{He}{i}, \ion{Ca}{ii} and \ion{Fe}{ii} lines.

\acknowledgments
We are grateful to the staff of the Subaru Telescope for their
kind support.
We thank Justyn Maund and the co-authors of 
Maund et al. (2007a) paper
for kindly providing their polarization data taken with VLT
(ESO VLT program 75.D-0213).
We are also grateful to the staff of VLT. 
M.T. and N.T. are supported by the JSPS 
(Japan Society for the Promotion of Science) 
Research Fellowship for Young Scientists.
This research has been supported in part by World Premier
International Research Center Initiative, MEXT,
Japan, and by the Grant-in-Aid for Scientific Research of the JSPS
(18104003, 18540231, 20540226) and MEXT (19047004, 20040004).


\begin{thebibliography}{}


\bibitem[Anupama et al.(2005)]{anu05} Anupama, G. C., Sahu, D. K., Deng, J., Nomoto, K., Tominaga, N., Tanaka, M., Mazzali, P. A., \& Prabhu, T. P. 2005,  631, 125

\bibitem[Blondin et al.(2003)]{blo03} Blondin, J.M., Mezzacappa, A., \& DeMarino, C., 2003, ApJ, 584, 971

\bibitem[Burrows et al.(1995)]{bur95} Burrows, A., Hayes, J., Fryxell, B. A. 1995, ApJ, 450, 830

\bibitem[Cropper et al.(1988)]{cro88} Cropper, M., Bailey, J., McCowage, J., Cannon, R. D., \& Couch, Warrick J. 1988, MNRAS, 231, 695

\bibitem[Davis \& Greenstein(1951)]{dav51} Davis, L. Jr., Greenstein, J. L. 1951, ApJ, 114, 206

\bibitem[Folatelli et al.(2006)]{fol06} Folatelli, G., et al. 2006, ApJ, 641, 1039

\bibitem[Gorosabel et al.(2006)]{gor06} Gorosabel, J., et al. 2006, A\&A, 459, L33

\bibitem[Harkness et al.(1987)]{har87} Harkness, R. P., et al. 1987, ApJ, 317, 355

\bibitem[Hoffman et al.(2008)]{hof08}Hoffman, J. L., Leonard, D. C., Chornock, R., Filippenko, A. V., Barth, A. J., Matheson, T. 2008, ApJ, 688, 1186

\bibitem[H\"oflich(1991)]{hof91} H\"oflich, P. 1991, A\&A, 246, 481

\bibitem[Hwang et al.(2004)]{hwa04} Hwang, U., et al. 2004, ApJ, 615, L117

\bibitem[Iwakami et al.(2008)]{iwa08} Iwakami, W., Kotake, K., Ohnishi, N., Yamada, S., \& Sawada, K 2008, ApJ, 678, 1207

\bibitem[Janka \& M\"uller(1996)]{jan96} Janka, H.-Th., M\"uller, E. 1996, A\&A, 306, 167

\bibitem[Kashikawa et al.(2002)]{kas02} Kashikawa, N., et al. 2002, PASJ, 54, 819

\bibitem[Kawabata et al.(2002)]{kaw02} Kawabata, K.S., et al. 2002, ApJ, 580, L39
\bibitem[Kawabata et al.(2003)]{kaw03} Kawabata, K.S., et al. 2003, ApJ, 593, L19

\bibitem[Khokhlov et al.(1999)]{kho99} Khokhlov, A. M., H\"oflich, P. A., Oran, E. S., Wheeler, J. C., Wang, L., \& Chtchelkanova, A. Yu. 1999, ApJ, 524, L107

\bibitem[Leonard et al.(2001)]{leo01} Leonard, D.C., Filippenko, A. V., Ardila, D. R., \&  Brotherton, M. S. 2001, ApJ, 553, 861

\bibitem[Leonard et al.(2002)]{leo02} Leonard, D.C., Filippenko, A.V., Chornock, R., \& Foley, R. 2002, PASP, 114, 1333

\bibitem[Leonard et al.(2006)]{leo06} Leonard, D.C., et al. 2006, Nature, 440, 505

\bibitem[Lucy(1991)]{luc91} Lucy, L. B. 1991, ApJ, 383, 308

\bibitem[Maeda et al.(2002)]{mae02} Maeda, K., et al. 2002, ApJ, 565, 405

\bibitem[Maeda et al.(2007)]{mae07} Maeda, K., et al. 2007, ApJ, 666, 1069

\bibitem[Maund et al.(2007a)]{mau07a} Maund, J. R., Wheeler, J. C., Patat, F., Baade, D., Wang, L., \& H\"oflich, P. 2007a, MNRAS, 381, 201 (M07)

\bibitem[Maund et al.(2007b)]{mau07b} Maund, J. R., Wheeler, J. C., Patat, F., Baade, D., Wang, L., \& H\"oflich, P. 2007b, A\&A, 475, L1

\bibitem[Maund et al.(2007c)]{mau07c} Maund, J. R., Wheeler, J. C., Patat, F., Wang, L., Baade, D., H\"oflich, P. A. 2007c, ApJ, 671, 1944

\bibitem[Mazzali et al.(2005)]{maz05} Mazzali, P.A. Benetti, S., Stehle, M., Branch, D., Deng, J., Maeda, K., Nomoto, K., \& Hamuy, M. 2005, MNRAS, 357, 200

\bibitem[McCall(1984)]{mcc84} McCall, M. L. 1984, MNRAS, 210, 829

\bibitem[Mikami et al.(2008)]{mik08} Mikami, H., Sato, Y., Matsumoto, T., \& Hanawa, T. 2008, ApJ, 683, 357

\bibitem[Modjaz et al.(2005)]{mod05} Modjaz, M., Kirshner, R.P., \& Challis, P. 2005, IAU Circ., 8522, 2

\bibitem[M\"uller \& Hillebrandt(1981)]{mue81} M\"uller, E., \& Hillebrandt, S. 1981, A\&A, 103, 358

\bibitem[Oke(1990)]{oke90} Oke, J.B. 1990, AJ, 99, 1621

\bibitem[Parrent et al.(2007)]{par07} Parrent, J., et al 2007, PASP, 119, 135

\bibitem[Patat \& Romaniello(2006)]{pat06} Patat, F. \& Romaniello, M., 2006, PASP, 118, 146

\bibitem[Scarrott et al. (1987)]{sca87} Scarrott, S. M., Ward-Thompson, D., \& Warren-Smith, R. F. 1987, MNRAS, 224, 299

\bibitem[Scarrott et al. (1993)]{sca93} Scarrott, S. M., Draper, P. W., Stockdale, D. P., \& Wolstencroft, R. D., 1993, MNRAS, 264, L7

\bibitem[Serkowski et al.(1975)]{ser75} Serkowski, K., Mathewson, D.L., \& Ford, V.L. 1975, ApJ, 196, 261

\bibitem[Schlegel et al.(1998)]{sch98} Schlegel, D. J., Finkbeiner, D. P., \& Davis, M. 1998, ApJ, 500, 525

\bibitem[Shapiro \& Sutherland(1982)]{sha82} Shapiro, P.R., \& Sutherland, P.G. 1982, ApJ, 263, 902

\bibitem[Takiwaki et al.(2004)]{tak04} Takiwaki, T., Kotake, K., Nagataki, S., \& Sato, K. 2004, ApJ, 616, 1086

\bibitem[Tanaka et al.(2008a)]{tan08a} Tanaka, M., Kawabata, K. S., Maeda, K., Hattori, T., \& Nomoto, K. 2008a, ApJ, 689, 1191 

\bibitem[Tanaka et al.(2008b)]{tan08b} Tanaka, M., et al. 2008b, ApJ, 677, 448

\bibitem[Tinbergen(1996)]{tin96} Tinbergen, J. 1996, Astronomical Polarimetry (New York: Cambridge Univ. Press)

\bibitem[Tominaga et al.(2005)]{tom05} Tominaga, N., et al. 2005, ApJ, 633, L97

\bibitem[Tominaga(2009)]{tom09} Tominaga, N., 2009, ApJ, 690, 526

\bibitem[Trammell et al.(1993)]{tra93} Trammell, S.R., Hines, D.C., \& Wheeler, J.C. 1993, ApJ, 414, L21

\bibitem[Turatto et al.(2003)]{tur03} Turatto M., Benetti S., Cappellaro E., 2003, in From Twilight to Highlight: The Physics of Supernovae, ed. by Hillebrandt W., Leibundgut B., Springer, Berlin, 200

\bibitem[Turnshek et al.(1990)]{tur90}	Turnshek, D. A., Bohlin, R. C., Williamson, R. L., II, Lupie, O. L., Koornneef, J., \& Morgan, D. H. 1990, AJ, 99, 1243

\bibitem[Wang et al.(2001)]{wan01} Wang, L., et al. 2001, ApJ, 550, 1030

\bibitem[Wang et al.(2002)]{wan02} Wang, L., et al. 2002, ApJ, 579, 671

\bibitem[Wang et al.(2003a)]{wan03a} Wang, L., Baade, D., H\"oflich, P., \& Whheler, J.C 2003, ApJ, 592, 457

\bibitem[Wang \& Baade(2005)]{wan05} Wang, L., \& Baade, D. 2005, IAU Circ., 8521, 2

\bibitem[Wang \& Wheeler(2008)]{wan08} Wang, L., \& Wheeler, J. C. 2008, ARA\&A, 46, 433

\bibitem[Whittet et al.(1992)]{whi92} Whittet, D. C. B., Martin, P. G., Hough, J. H., Rouse, M. F., Bailey, J. A., \& Axon, D. J. 1992, ApJ, 386, 562

\bibitem[Yamada \& Sato(1994)]{yam94} Yamada, S., \& Sato, K. 1994, ApJ, 434, 268

\end{thebibliography}
\end{document}